\documentclass{article}
\usepackage{spconf,amsmath,graphicx,amssymb}
\usepackage{booktabs}
\usepackage{diagbox}
\usepackage{multirow}
\usepackage{color}
\usepackage{subfigure}
\usepackage{hyperref}
\usepackage{float}
\usepackage{url}

\title{Multi-GradSpeech: Towards Diffusion-based Multi-Speaker Text-to-speech Using Consistent Diffusion Models}
\name{Heyang Xue, Shuai Guo, Pengcheng Zhu, Mengxiao Bi}
\address{Fuxi AI Lab, NetEase Inc., Hangzhou, China}

\begin{document}
%
\maketitle
\begin{abstract}
Despite imperfect score-matching causing drift in training and sampling distributions of diffusion models, recent advances in diffusion-based acoustic models have revolutionized data-sufficient single-speaker Text-to-Speech~(TTS) approaches, with Grad-TTS being a prime example. However, the sampling drift problem leads to these approaches struggling in multi-speaker scenarios in practice due to more complex target data distribution compared to single-speaker scenarios. In this paper, we present Multi-GradSpeech, a multi-speaker diffusion-based acoustic models which introduces the Consistent Diffusion Model~(CDM) as a generative modeling approach. We enforce the consistency property of CDM during the training process to alleviate the sampling drift problem in the inference stage, resulting in significant improvements in multi-speaker TTS performance. Our experimental results corroborate that our proposed approach can improve the performance of different speakers involved in multi-speaker TTS compared to Grad-TTS, even outperforming the fine-tuning approach. Audio samples are available at \url{https://welkinyang.github.io/multi-gradspeech/}
\end{abstract}

\vspace{-10pt}
\section{Introduction}
\label{sec:intro}
\vspace{-5pt}
Text-to-Speech (TTS) systems that use deep learning rely on datasets with one or more speakers to train models that can produce high-quality speech. This approach consists of two key components: an acoustic model that converts text into frame-level acoustic features, such as mel-spectrograms, and a vocoder that converts these features into waveforms. The acoustic model plays an essential role in controlling factors such as prosody, speaker identity, and style, ultimately contributing to the naturalness and expressiveness of generated speech. While it is straightforward to train a single-speaker acoustic model to generate the target speaker's voice, supporting multiple speakers is critical to creating an economical and efficient multi-speaker TTS system or as pre-trained models for fine-tuning to improve the performance of the generated speech. On the other hand, improving multi-speaker acoustic modeling is a widely researched topic due to the greater complexity of target data distribution in multi-speaker scenarios compared to single-speaker scenarios.

Researchers have long sought to investigate generative modeling approaches~\cite{van2016pixel, DBLP:conf/nips/GoodfellowPMXWOCB14, DBLP:conf/icml/RezendeM15, DBLP:conf/nips/HoJA20} to improve multi-speaker acoustic modeling. Auto-regressive models, which include RNNs or Transformers, have been used as generative models for acoustic models~\cite{DBLP:conf/interspeech/WangSSWWJYXCBLA17, DBLP:conf/interspeech/ChenTRXSZQ20, yu20c_interspeech} . However, these models suffer from issues such as accumulated error and exposure bias~\cite{DBLP:conf/interspeech/GuoSHX19, DBLP:conf/icassp/0008SLBG020} during the inference phase, which can reduce the quality of multi-speaker acoustic modeling. Beyond this, some studiess~\cite{DBLP:conf/icassp/ChienLHHL21, DBLP:conf/icassp/LiOLH21, DBLP:conf/slt/LiOLH21} have used non-autoregressive models optimized by L1 or L2 losses as acoustic models for multi-speaker acoustic modeling. However, these models often produce over-smoothed speech, resulting from the assumption that data noise is drawn from Laplace or zero-mean Gaussian distribution~\cite{DBLP:journals/access/HeraviH18}, which may not be accurate. In recent years, generative adversarial networks~(GANs) and normalizing flows~(NFs) have been applied as generative modeling approaches to address the over-smoothing problems in multi-speaker acoustic modeling~\cite{DBLP:conf/interspeech/YangBBKC21, DBLP:conf/nips/KimKKY20}.

In recent years, diffusion models have emerged as a potent class of generative networks, particularly in the domain of image generation. These models have also been applied to acoustic modeling, producing impressive speech synthesis quality, as reported in recent researches~\cite{DBLP:conf/icml/PopovVGSK21, jeong21_interspeech, liu2022diffgan}. However, current efforts primarily focus on single-speaker scenarios and the sampling drift problem of the diffusion models~\cite{DBLP:journals/corr/abs-2302-09057} have locked the full potential of these approaches for multi-speaker TTS scenarios due to more complex target data distribution. Specifically, these approaches suffer from quality degradation from single-speaker scenarios to multi-speaker scenarios when there is ample data available for the target speakers. Furthermore, over-smoothing persists in the generated results when target speaker data is scarce. 

This paper proposes Multi-GradSpeech, a novel solution to address the challenges mentioned earlier. Multi-GradSpeech utilizes the Consistent Diffusion Model (CDM)~\cite{DBLP:journals/corr/abs-2302-09057} for generative modeling, which tackles the sampling drift problem during the inference phase to enhance the performance of diffusion-based multi-speaker acoustic modeling. The experiments conducted on a Mandarin multi-speaker dataset demonstrated that Multi-GradSpeech performs better than Grad-TTS~\cite{DBLP:conf/icml/PopovVGSK21} for diffusion-based multi-speaker modeling without the need for fine-tuning.

\begin{figure}[t]
    \centering
    \includegraphics[width=8.5cm,height=9.5cm]{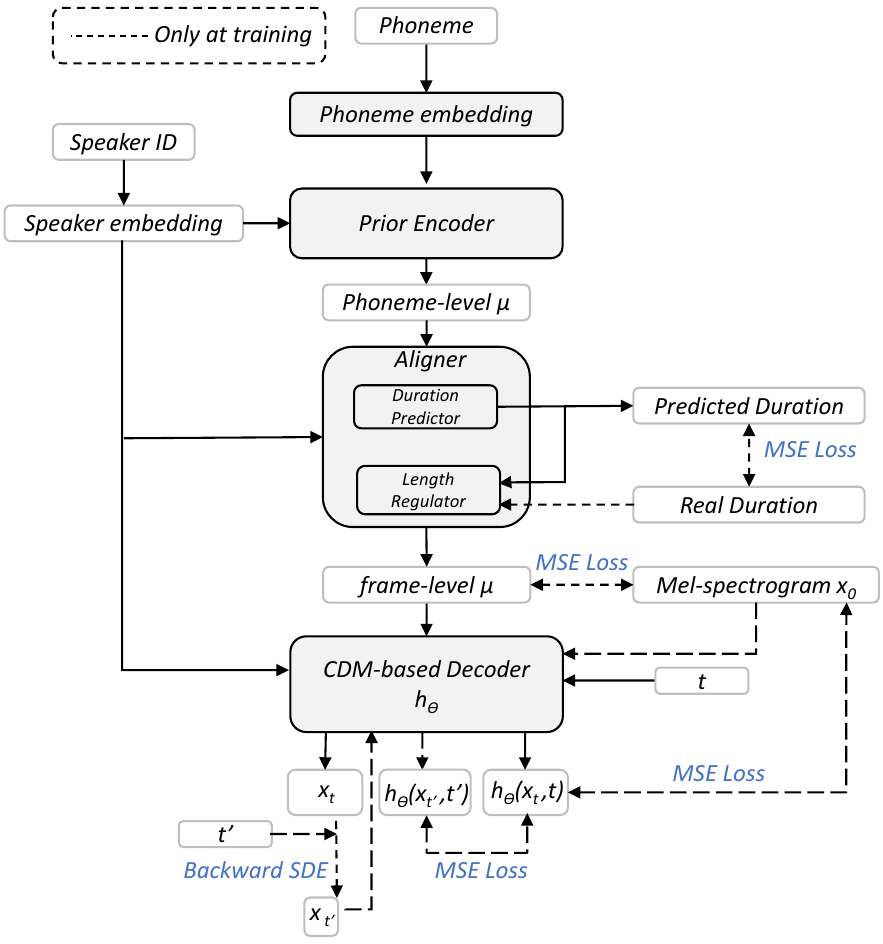}
    \caption{The overall architecture of Multi-GradSpeech, where the dashed line appears only in training.}
    \label{fig:multi-gradspeech}
    \vspace{-15pt}
\end{figure}

\vspace{-10pt}
\section{Background}
\label{sec:background}
\vspace{-10pt}
Basically diffusion models are a class of generative models that gradually add noise to real data through a forward process. While the reverse process which can be approximated using a neural network removes noise from a simple distribution. In the context of stochastic differential equations~(SDE)~\cite{DBLP:conf/iclr/0011SKKEP21}, defining $p_0$ as the data distribution, the forward and backward processes can be driven by two SDEs respectively. The forward SDE is defined as:
\vspace{-5pt}
\begin{equation}
d x_t=g(t) d B_t, x_0 \sim p_0, x_t \sim N\left(x_0, \sigma_t^2 I_d\right).
\end{equation}
and the backward SDE is defined as:
\begin{equation}
d x_t=-g(t)^2 \nabla_x \log p\left(x_t, t\right) d t+g(t) d \bar{B}_t
\label{b_sde}
\end{equation}
where $t \in[0, T]$, $\sigma_t$ is the noise schedule of the diffusion process, $B_t$ is a Brownian motion, $g(t)^2 = \frac{d\sigma_t^2}{dt}$ and $p\left(x_t, t\right)$ is the probability density of $x_t$.
It is worth noting that for all backward SDEs there exists a corresponding deterministic process satisfying the following Ordinary Differential Equation~(ODE):
\vspace{-7pt}
\begin{equation}
\label{b_ode}
d x_t= -\frac{1}{2}g(t)^2\nabla_x \log p\left(x_t, t\right).
\vspace{-5pt}
\end{equation}
Since this ODE and the corresponding backward SDE share the identical  marginal probability density, we can generate samples by solving the ODE using first-order Euler solver or other higher-order solvers. However, for both backward SDE and ODE, $\nabla_x \log p\left(x_t, t\right)$ which is also known as the score function is intractable. Several studies proposed to approximate it by a time-conditional denoiser that is to say predicting the true data $x_0$ from corrupted data $x_t$. Specifically, we aim to use neural networks to learn the function $h$: $\mathbb{R}^d \times[0,1] \rightarrow \mathbb{R}^d$:
\begin{equation}
h(x, t)=\mathbb{E}\left[x_0 \mid x_t=x\right]
\end{equation}
where the expectation is over $x_0$ by running the backward process with the initial condition $x_{t} = x$. By leveraging  Tweedie's formula, we can establish the relationship between the score function and the denoiser $h$:
\begin{equation}
\label{tweedie_formula}
\nabla_x \log p(x, t)=\frac{h(x, t)-x}{\sigma_t^2}
\end{equation}
Bringing  Eq.~(\ref{tweedie_formula}) into  Eq.~(\ref{b_sde}) and  Eq.~(\ref{b_ode}) we can use the trained denoiser for sampling. Denoising score matching loss is commonly used to train h:
\begin{equation}
\label{loss_dsm}
L_{DSM}=\mathbb{E}_{x_0 \sim p_0, x_t \sim N\left(x_0, \sigma_t^2 I_d\right)}\left\|h_\theta\left(x_t, t\right)-x_0\right\|^2,
\end{equation}
where $\theta$ is a neural network.

Perfectly learning the real denoiser $h^{*}$ is commonly unattainable, which leads to the sampling drift challenge. Specifically, during the sampling stage $x_t$ is derived from an imperfect $h^{*}_{\theta}$, and each time step produces a drift with respect to the true distribution $p^{*}_{t}$, which ultimately leads to the generation of samples far from the true distribution due to error accumulation. To mitigate sampling drift, Daras et al. proposed Consistent Diffusion Model (CDM). In \cite{DBLP:journals/corr/abs-2302-09057}, a denoiser function h is said to be consistent if for all $t \in[0, T]$ and all $x \in \mathbb{R}^d$, 
\begin{equation}
h(x, t)=\mathbb{E}_h\left[x_0 \mid x_t=x\right]
\end{equation}
where the expectation is over $x_t^{\prime}$ by solving Eq.~(\ref{b_sde}) or Eq.~(\ref{b_ode}) starting at $x_t$ with function $h$. Further, a consistency loss is added to the original loss for enforcing consistency property of the denoiser:
\begin{equation}
\label{loss_cdm}
\resizebox{.98\hsize}{!}{
$
 L_{CDM}=\mathbb{E}_{x_t \sim p_t} \mathbb{E}_{t^{\prime} \sim \mathcal{U}[t-\epsilon,
 t]}\Bigg[\mathbb{E}_\theta\left[h_\theta\left(x_{t^{\prime}}, t^{\prime}\right) \mid x_t=x\right]-h_\theta(x, t)\Bigg]^2 / 2
$
}
\end{equation}
where the innermost expectation is over $x_{t^{\prime}}$. To reduce the computation time of this loss function, $t^{\prime}$ is generally chosen to be close to $t$ and the backward process from $x_t$ to $x_t{\prime}$ is discretized into a small number of steps.

\vspace{-8pt}
\section{Multi-GradSpeech}
\vspace{-8pt}
\label{sec:pagestyle}
Although diffusion-based acoustic models, such as Grad-TTS, show good results in single-speaker TTS when adequate data is available, their performance in multi-speaker TTS still has room for improvement. In this study, we propose a CDM-based multi-speaker acoustic modeling approach, Multi-GradSpeech, to enhance the performance of multi-speaker acoustic models. The proposed approach comprises three modules: a prior encoder, an aligner, and a CDM-based decoder (shown in Fig.~\ref{fig:multi-gradspeech}). We describe Multi-GradSpeech's training and inference in detail. 
\vspace{-10pt}
\subsection{Training}
\vspace{-5pt}
During the training phase, Multi-GradSpeech is optimized using prior loss, duration loss, and diffusion-based loss. The architecture of Multi-GradSpeech is illustrated in Fig. \ref{fig:multi-gradspeech}. Multi-GradSpeech takes phoneme and speaker id as inputs and encodes them with phoneme and speaker embeddings, respectively. The encoded inputs are fed into the prior encoder, which generates phoneme-level intermediate representations with the same number of channels as the real mel-spectrogram~(denoted as $\mu$). The phoneme-level $\mu$ and speaker embedding are then fed into the aligner to predict the duration and frame-level $\mu$, and compute the mean square error~(MSE) loss between predicted and real durations and mel-spectrogram~(denoted as $x_{0}$), respectively.

Next, we feed the frame-level $\mu$, time steps~(denoted as $t$), and speaker embedding into a CDM-based decoder~(denoted as $h_{\theta}$) to generate $h_{\theta}(x_{t}, t)$, where $condition$ is ignored for simplicity. The denoiser $h_{\theta}$ is trained by computing the MSE loss between $x_{0}$ and $h_{\theta}(x_{t}, t)$, which is referred to as $L_{DSM}$ and mentioned in Eq.~\ref{loss_dsm}.

While denoising score matching loss $L_{DSM}$ can optimize  $h_{\theta}$, the challenge of sample drift in the inference phase degrades the performance of diffusion-based acoustic models in multi-speaker scenarios. Therefore, we further incorporate the consistency loss function also referred to as $L_{CDM}$ in Eq.~\ref{loss_cdm} to enhance the consistency property of $h_{\theta}$. Specifically, we first sample $t^{'}$ where $t^{'} \sim uniform(t-\epsilon, t)$ and $\epsilon$ determines how close $t^{'}$ is to $t$. Then, a first-order backward SDE is performed for several time steps with $t$ as the starting point and $t^{'}$ as the end point to obtain $x_{t^{'}}$. Finally, $x_{t^{'}}$ and $t^{'}$ are fed into $h_{\theta}$ to obtain $h_{\theta}(x_{t^{'}}, t^{'})$ and calculate the MSE loss with $h_{\theta}(x_{t}, t)$. Thus, the final loss for Multi-GradSpeech is as follows, where $\lambda$ controls the scale of $L_{CDM}$:
\begin{equation}
    L_{final} = L_{duration} + L_{prior} + L_{DSM} + \lambda\mathbb{E}_{t}\left[L_{CDM}\right]
\end{equation}

\vspace{-15pt}
\subsection{Inference}
\vspace{-5pt}
Similar to the training phase, Multi-GradSpeech first predicts the frame-level $\mu$ except that the duration is predicted by the duration predictor instead of using real duration. Then $x_{0}$ is progressively recovered by solving the ODE in E.q~\ref{b_ode} using the stochastic solver proposed in \cite{Karras2022edm}. Note that at this point $\nabla_x \log p\left(x_t, t\right)$ in E.q~\ref{b_ode} is approximated by $h_{\theta}\left(x_t, t\right)$ through E.q~\ref{tweedie_formula}.

\vspace{-15pt}

\section{Experiments}
\label{sec:exp}
\vspace{-10pt}
\subsection{Data Setups}
\vspace{-5pt}
In our study, we conducted experiments on a Mandarin Chinese dataset featuring multiple speakers. This dataset was comprised of 50 Mandarin Chinese speakers, each with between 500 to 10,000 paired audio and transcript files, adding up to a total of 123.73 hours. The original sampling rate of the data was 48Khz and we resampled it to 16Khz. We extracted mel-spectrograms utilizing a hop size of 200 and a window size of 800 for acoustic modeling. We used the Montreal Forced Aligner to obtain the phoneme durations.

\vspace{-10pt}
\subsection{Implementation Details}
\vspace{-5pt}
The Multi-GradSpeech system utilizes the same architectural configurations for the prior encoder, duration predictor, and CDM-based decoder as seen in Grad-TTS. Solving the first-order backward SDE from time $t$ to $t^{'}$ is necessary to compute $L_{CDM}$. In accordance with~\cite{DBLP:journals/corr/abs-2302-09057}, we take six steps in the backward SDE with $\epsilon$ set to 0.05 and a lambda value of 2 to impose consistency constraints. In the inference phase, we set parameters $S_{churn}$, $S_{min}$, $S_{max}$, and $S_{noise}$ to 11, 0.05, 15, and 1.003 respectively for the stochastic sampler and take 18 backward steps. We use the modified RefineGAN proposed in \cite{DBLP:conf/interspeech/XueWZXZB22} as the vocoder except that we modified the parameters of the up-sampling network to make the output with a sample rate of 48Khz. 

\vspace{-10pt}

\subsection{Evaluation}
\vspace{-5pt}
\subsubsection{Target speakers}
\vspace{-5pt}
In order to evaluate the effectiveness of Multi-GradSpeech, we carefully selected two sets of speakers that vary greatly in the amount of data available for evaluation. Each set includes one male and one female speaker to ensure gender balance. The first set, which we refer to as Speakers-S, has sufficient data for single-speaker acoustic modeling~(9.09 hours and 10.93 hours). Conversely, the second set Speakers-I, has 1.47 hours and 1.13 hours of data, which is generally considered insufficient for single-speaker acoustic modeling. Our aim is to conduct a comparative analysis of Multi-GradSpeech's performance on these two sets of speakers. 

\vspace{-10pt}
\subsubsection{Metrics and comparison approaches}
\vspace{-5pt}
We conducted subjective and objective evaluations to measure the quality of samples generated by Multi-GradSpeech as compared to ground truth recordings and Grad-TTS. The subjective evaluation consisted of a Mean Opinion Score~(MOS) and Similarity MOS~(SMOS) test, while the objective evaluation used Word Error Rate~(WER) as a metric. 
\begin{table*}[htbp]
  \centering
  \caption{Subjective and objective evaluations on different speakers.}
    \resizebox*{14cm}{4.375cm}{
    \begin{tabular}{cccccccc}
    \toprule
    \multirow{3}[2]{*}{\textbf{Model}} & \multirow{3}[2]{*}{\textbf{Training method}} & \multicolumn{6}{c}{\textbf{Target Speaker}} \\
          &       & \multicolumn{3}{c}{\textbf{Speakers-S}} & \multicolumn{3}{c}{\textbf{Speakers-I}} \\
          &       & MOS (↑) & SMOS (↑) & WER (↓) & MOS (↑) & SMOS (↑) & WER (↓) \\
    \midrule
    \multirow{3}[2]{*}{\textbf{Grad-TTS}} & Single-speaker & 3.50±0.13 & 4.20±0.02 & 5.13 & 2.13±0.12 & 3.47±0.06 & 27.80 \\
          & Multi-speaker & 3.35±0.12 & 4.16±0.02 & 5.14 & 2.73±0.12 & 3.80±0.05  & 8.94 \\
          & Finetune & 3.43±0.13 & 4.19±0.02 & 6.02 & 2.80±0.11 & 3.84±0.05 & 9.07 \\
    \midrule
    \multirow{3}[2]{*}{\textbf{Multi-GradSpeech}} & Single-speaker & \textbf{3.92±0.12} & \textbf{4.22±0.02} & 3.99 & 2.80±0.13 & 3.69±0.06 & 17.85 \\
          & Multi-speaker & 3.90±0.13 & 4.19±0.02 & \textbf{3.10} & \textbf{3.23±0.13} & \textbf{3.82±0.05} & \textbf{6.06} \\
          & Finetune & 3.91±0.12 & 4.21±0.02 & 3.75 & 2.98±0.14 & 3.74±0.05 & 8.57 \\
    \midrule
    \textbf{Multi-GradSpeech (w/o $L_{CDM}$)} & Multi-speaker & 3.88±0.12 & 4.18±0.02  & 3.34  & 2.94±0.14 & 3.72±0.06 & 6.57 \\
    \midrule
    \multirow{2}[2]{*}{\textbf{Recording}} & \multirow{2}[2]{*}{} & \multicolumn{3}{c}{MOS (↑)} & \multicolumn{3}{c}{WER (↓)} \\
          &       & \multicolumn{3}{c}{4.66±0.12} & \multicolumn{3}{c}{1.72} \\
    \bottomrule
    \end{tabular}%
    }
  \label{tab:exp}%
  \vspace{-15pt}
\end{table*}%

To evaluate MOS, we selected 20 sentences from the open-source Chinese Standard Mandarin Speech Corpus that were not part of the training data. We generated results for 15 listeners to score on a scale of 1 to 5. For the WER evaluation, we employed a pre-trained ASR model from the FunASR framework~\cite{gao2023funasr}. To produce a multi-speaker version of Grad-TTS, we added speaker embedding to its prior encoder, duration predictor, and decoder to control speaker identity. Additionally, we used the Maximum Likelihood SDE solver and took 20 steps for discretization for Grad-TTS. The experimental results are shown in Table~\ref{tab:exp}. 

\begin{figure}
    \centering
    \includegraphics[width=1.0\linewidth]{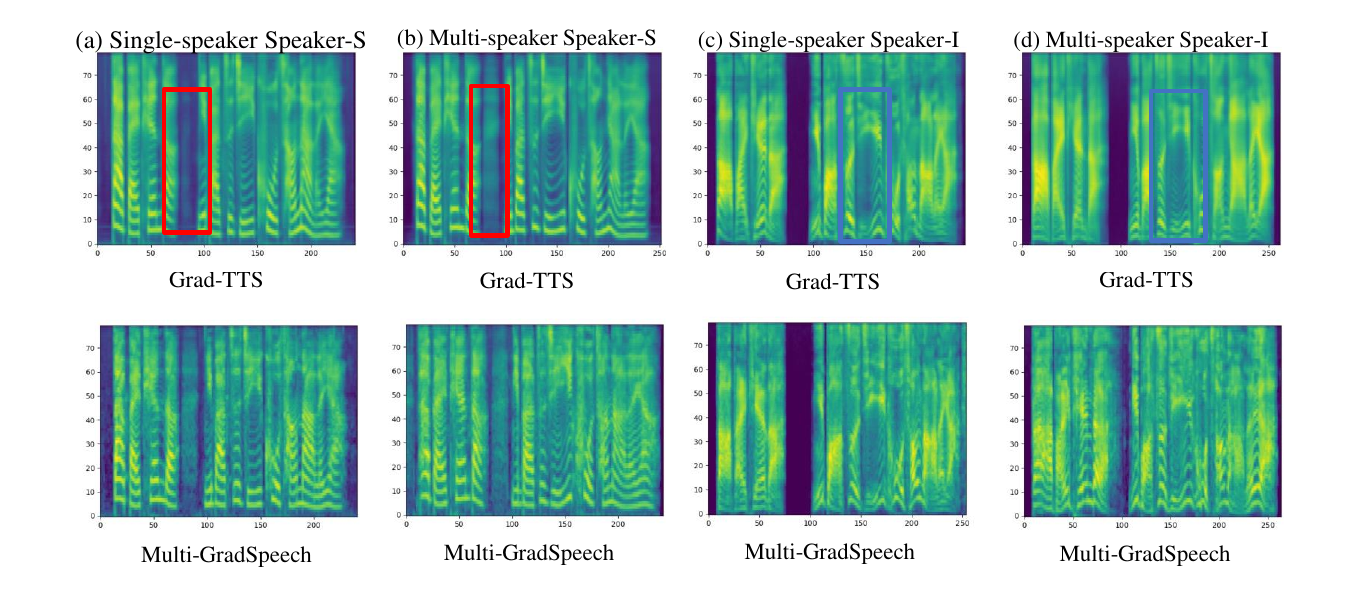}
     \vspace{-0.1in}
     \vspace{-15pt}
    \caption{Visualizations of generated mel-spectrograms by different approaches.}
    \vspace{-0.15in}
    \label{fig:mel_comparison}
    \vspace{-10pt}
\end{figure}

\vspace{-12pt}
\subsubsection{Analysis of subjective and objective results}
\vspace{-5pt}
In this study, we first trained single-speaker Grad-TTS and Multi-GradSpeech models using data from Speakers-I and Speakers-S. The single-speaker versions of both models demonstrated strong performance on both subjective and objective metrics when tested on Speakers-S. However, we observed a significant drop in performance when testing on Speakers-I, suggesting that the amount of data available for training an individual speaker model was insufficient. The Multi-GradSpeech model outperformed Grad-TTS on all metrics, indicating its superior modeling capabilities.

We then trained multi-speaker versions of both models using the whole dataset and observed the performance on two sets of speakers. We found that while Grad-TTS showed a slight decrease in MOS value and no change in WER, Multi-GradSpeech maintained its performance for Speakers-s and saw a significant decrease in WER. Moreover, both models showed significant improvements in MOS values and WER for Speakers-I, indicating the benefits of multi-speaker modeling.

We also fine-tuned the multi-speaker models for the four target speakers and the parameters of prior encoder were frozen and the other parts were updated, but observed no improvements in performance. While this may be due to the need for careful parameter design and fine-tuning epochs, it suggests that Multi-GradSpeech can build a strong multi-speaker TTS without fine-tuning. Notably, all versions of both models demonstrated high speaker similarity performance on Speakers-S, although SMOS values were reduced on Speakers-I.

Finally, we trained a version of Multi-GradSpeech without the CDM component~(remove $L_{cdm}$) to demonstrate its importance. The resulting model exhibited noticeable performance degradation across all metrics, highlighting the crucial role of CDM in multi-speaker TTS.

\vspace{-10pt}
\subsubsection{Analysis of visualization results}
\vspace{-5pt}
In order to uncover the reasons for the disparate performance of certain approaches in subjective and objective experiments, we examined mel-spectrograms generated by these approaches in a test case which is shown in Fig .\ref{fig:mel_comparison}. Specifically, we selected individual speakers from the Speakers-S and Speakers-I, which we denoted as Speaker-S and Speaker-I. We synthesized this test case using both single-speaker and multi-speaker versions of Grad-TTS and Multi-GradSpeech.  Initially, we compared the results of Grad-TTS generation on Speaker-S, noting that there was no significant difference betwee single-speaker and multi-speaker versions, with the exception of a small section in a red box. We discovered that the silence within this part had been over-smoothed in the multi-speaker version, giving it a metallic and unnatural sound. This helps explain why the WER did not decrease when Grad-TTS switched from the single-speaker version to the multi-speaker version on Speakers-S, but the MOS value decreased.

By contrast, when we applied Grad-TTS to Speaker-I and compared the single-speaker and multi-speaker versions, we found that the multi-speaker version did lead to more complete articulation. However, the lack of spectral detail in the overall sound persisted, leading to an over-smoothed result. This discrepancy explains why there was a significant decrease in WER when Grad-TTS transitioned from the single-speaker version to the multi-speaker version on Speakers-S, but the MOS value is only about 2.7.

Finally, when considering the single-speaker and multi-speaker versions of Multi-GradSpeech on Speaker-S, no significant differences were observed. Furthermore, we found that the over-smoothing problem did not appear in the multi-speaker version of Multi-GradSpeech on Speaker-I, and adequate detail was present in the middle and high frequencies.

\vspace{-10pt}
\section{CONCLUSION AND Future WORK}
\vspace{-5pt}
We introduce Multi-GradSpeech, a diffusion-based acoustic model for multi-speaker modeling. By utilizing Consistent Diffusion Models, we demonstrate significant improvements in our approach compared to Grad-TTS. We anticipate that Multi-GradSpeech will serve as an alternative to prior diffusion-based acoustic models, particularly in large TTS.

 \clearpage
\bibliographystyle{IEEE}
\bibliography{refs}

\end{document}